\documentclass[proof]{pasj01}
\Received{$\langle$reception date$\rangle$}
\Accepted{$\langle$acception date$\rangle$}
\Published{$\langle$publication date$\rangle$}

\begin{document}

\title{CO observations of the molecular gas in the Galactic H{\sc ii} region Sh2-48; Evidence for cloud-cloud collision as a trigger of high-mass star formation}
\author{Kazufumi Torii\altaffilmark{1}, Yusuke Hattori\altaffilmark{2}, Mitsuhiro Matsuo\altaffilmark{1}, Shinji Fujita\altaffilmark{2}, Atsushi Nishimura\altaffilmark{2}, Mikito Kohno\altaffilmark{2}, Mika Kuriki\altaffilmark{3}, Yuya Tsuda\altaffilmark{5}, Tetsuhiro Minamidani\altaffilmark{1,6}, Tomofumi Umemoto\altaffilmark{1,6}, Nario Kuno\altaffilmark{3,4}, Satoshi Yoshiike\altaffilmark{2}, Akio Ohama\altaffilmark{2}, Kengo Tachihara\altaffilmark{2}, Yasuo Fukui\altaffilmark{2}, Kazuhiro Shima\altaffilmark{7}, Asao Habe\altaffilmark{7}, Thomas J. Haworth\altaffilmark{8}}%
\altaffiltext{1}{Nobeyama Radio Observatory, 462-2 Nobeyama Minamimaki-mura, Minamisaku-gun, Nagano 384-1305, Japan}
\altaffiltext{2}{Graduate School of Science, Nagoya University, Chikusa-ku, Nagoya, Aichi 464-8601, Japan}
\altaffiltext{3}{Department of Physics, Graduate School of Pure and Applied Sciences, University of Tsukuba, 1-1-1 Ten-nodai, tsukuba, Ibaraki 305-8577, Japan}
\altaffiltext{4}{Tomonaga Center for the History of the Universe, University of Tsukuba, Tsukuba, Ibaraki 305-8571, JAPAN}
\altaffiltext{5}{Meisei University, 2-1-1 Hodokubo, Hino, Tokyo 191-0042, Japan}
\altaffiltext{6}{Department of Astronomical Science, School of Physical Science, SOKENDAI (The Graduate University for Advanced Studies), 2-21-1, Osawa, Mitaka, Tokyo 181-8588, Japan}
\altaffiltext{7}{Faculty of Science, Department of Physics, Hokkaido University, Kita 10 Nishi 8 Kita-ku, Sapporo 060-0810, Japan}
\altaffiltext{8}{Astrophysics Group, Imperial College London, Blackett Laboratory, Prince Consort Road, London SW7 2AZ, UK}

\email{kazufumi.torii@nao.ac.jp}


\maketitle

\begin{abstract}
Sh2-48 is a Galactic H{\sc ii} region located at 3.8\,kpc with an O9.5-type star identified at its center.
As a part of the FOREST Unbiased Galactic plane Imaging survey using the Nobeyama 45-m telescope (FUGIN) project, we obtained the CO $J$=1--0 dataset for a large area of Sh2-48 at a spatial resolution of 21$''$ ($\sim$0.4\,pc), which we used to find a molecular cloud with a total molecular mass of $\sim3.8\times10^4$\,$M_\odot$ associated with Sh2-48.
The molecular cloud has a systematic velocity shift within a velocity range $\sim$42--47\,km\,s$^{-1}$.
On the lower velocity side the CO emission spatially corresponds with the bright 8\,$\mu$m filament at the western rim of Sh2-48, while the CO emission at higher velocities is separated at the eastern and western sides of the 8\,$\mu$m filament.
This velocity change forms V-shaped, east-west-oriented feature on the position-velocity diagram.
We found that these lower and higher-velocity components are, unlike the infrared and radio continuum data, physically associated with Sh2-48.
To interpret the observed V-shaped velocity distribution, we assessed a cloud-cloud collision scenario and found from a comparison between the observations and simulations that the velocity distribution is an expected outcome of a collision between a cylindrical cloud and a spherical cloud, with the cylindrical cloud corresponding to the lower-velocity component, and the two separated components in the higher-velocity part interpretable as the collision-broken remnants of the spherical cloud.
Based on the consistency of the $\sim$1.3\,Myr estimated formation timescale of the H{\sc ii} region with that of the collision, we concluded that the high-mass star formation in Sh2-48 was triggered by the collision.
\end{abstract}

\section{Introduction}
High-mass stars exert enormous influence over the galactic environment via injecting large amounts of energy by stellar winds, ultra-violet (UV) radiation, and supernova explosions. 
It is therefore crucial to our long-term goal of elucidating the pattern of galactic evolution to understand the formation mechanism of high-mass stars.

Recent attempts to search for parental molecular clouds of high-mass stars have shown that super-sonic collision of molecular clouds is a promising mechanism for triggering the formation of such stars (e.g., \cite{lor1976, fur2009, tor2011, fuk2014, fuk2017b}).
These observations have been carried out toward the high-mass star-forming regions in the Milky Way and the Large Magellanic Cloud, which include H{\sc ii} regions excited by single O stars (e.g., M20: \cite{tor2011,tor2017} and RCW120: \cite{tor2015}), super star clusters with a few tens O stars within a small volume (e.g., Westerlund\,2: \cite{fur2009,oha2010}, NGC3603: \cite{fuk2014}, and RCW38: \cite{fuk2016}), large H{\sc ii} regions extending for several tens of pc (e.g., W51: \cite{oku2001,fuj2017ccc1} and NGC6334+NGC6357: \cite{fuk2017ccc2}), and so on.
Extensive theoretical studies of cloud-cloud collision (CCC) have also been performed (e.g., \cite{sto1970a, hab1992, ino2013, tak2014, haw2015b}).
The results of these studies indicate the importance of CCC in high-mass star formation, although observational samples remain limited. 
It is therefore important to increase the sampling of CCC regions to obtain a more comprehensive understanding of CCC and the subsequent high-mass star formation.

Important targets for the observational study of CCC are infrared ring or bubble structures distributed in the Galactic plane, which were extensively studied by \citet{chu2006} based on {\it Spitzer} observations.
The authors identified about 600 ring-like 8\,$\mu$m structures within $\pm60^\circ$ in the Galactic longitude, followed by an expanded catalog of $\sim$5,100 ring-like emissions by \citet{sim2012}. 
As indicated by \citet{deh2010} and \citet{ken2012}, many of the identified infrared ring-like structures enclose H{\sc ii} regions. 
It is therefore important in investigating the CCC model as the mechanism of high-mass star formation to analyze the molecular line data in these infrared ring-like structures. 
To this end, in this study we focused on the Galactic H{\sc ii} region Sh2-48 which was cataloged as N18 by \citet{chu2006}.

\begin{figure}
 \begin{center}
  \includegraphics[width=16cm]{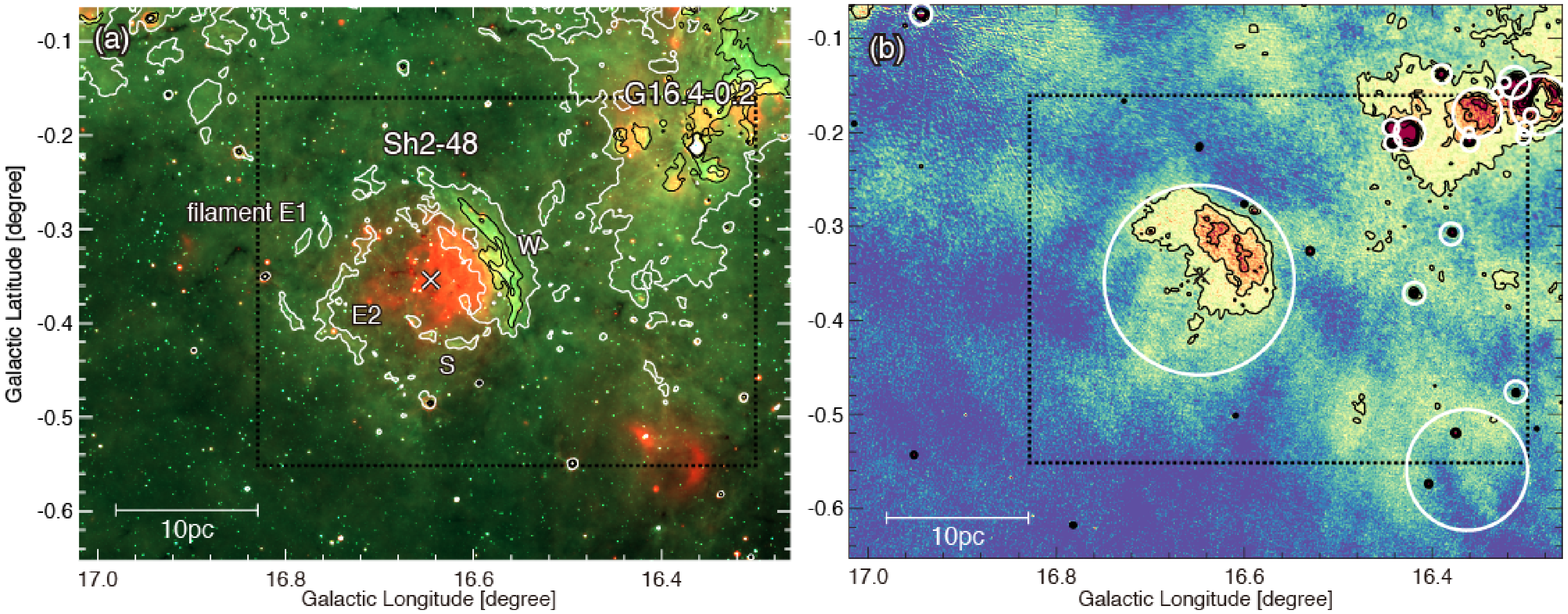}
 \end{center}
 \caption{Composite of the {\it Spizter}/MIPSGAL 24\,$\mu$m (red) and {\it Spizter}/IRAC 8\,$\mu$m (green) images toward a large area of Sh2-48 \citep{ben2003, chu2009, car2009}. Contours indicate the {\it Spizter}/IRAC 8\,$\mu$m emission, to which a median filter was applied to the data with a window size of $20'' \times 20''$, and are plotted at 5, 7, 9, and 11\,$\sigma$ of the local background noise level, which correspond to 90, 120, 150, and 180\,MJy\,str$^{-1}$, respectively. 
 The lowest contours are shown in white, while the others are in black.
 (b) Image and contours show the MAGPIS 20\,cm radio continuum emission obtained using VLA \citep{hel2006}. The white circles indicate the H{\sc ii} regions listed in the WISE galactic H{\sc ii} region catalog \citep{and2014}. Contours start at 5$\sigma$ and repeats in intervals of 2$\sigma$ of the local background noise level, which correspond to 0.48 and 1.2\,mJy\,beam$^{-1}$, respectively. In (a) and (b) the cross depicts BD-14\,5014 (O9.5V star), while the black box with dashed lines indicate the region in which the FUGIN data were analyzed. }\label{fig:rgb}
\end{figure}

Sh2-48 is a Galactic H{\sc ii} region located at $(l,b)\sim(16\fdg65, -0\fdg35)$, that was first cataloged by \citet{sha1959}.
Figure\,\ref{fig:rgb}(a) shows a composite of the {\it Spizter}/MIPSGAL 24\,$\mu$m (red) and {\it Spizter}/IRAC 8\,$\mu$m (green) images toward a large area of Sh2-48 \citep{ben2003, par2005, chu2009, car2009}, with the H{\sc ii} regions cataloged by the Wide-field Infrared Survey Explorer (WISE) observations depicted by white circles \citep{and2014}. 
Figure\,\ref{fig:rgb}(b) shows the MAGPIS interferometric radio continuum data at 20\,cm obtained using the Very Large Array (VLA) \citep{hel2006}. 
The white circles depict the H{\sc ii} regions listed in the WISE Galactic H{\sc ii} region catalog \citep{and2014}.
A star (BD-14\,5014) of spectral type O9.5V has been identified at the center of Sh2-48 \citep{ave1984, vog1975, vij1993}.
The 8\,$\mu$m emission in Sh2-48, which is attributed to polycyclic aromatic hydrocarbon (PAH) excitation by UV radiation, primarily comprises four filamentary structures, dubbed filaments W, S, E1, and E2 in Figure\,\ref{fig:rgb}, that form a bubbling structure with a size of $\sim$10$'$.
These 8\,$\mu$m filaments enclose the 20\,cm emission from the ionized gas and the 24\,$\mu$m emission from warm dust grains that are being heated by the H{\sc ii} region.

\citet{loc1989} detected radio recombination lines in Sh2-48 at a radial velocity of $\sim44.9$\,km\,s$^{-1}$, which corresponds to near and far distances of 3.8\,kpc and 12.4\,kpc, respectively, under the assumption of a flat Galactic disk rotational model with the solar galactocentric distance $R_\odot$ of $7.6\pm0.3$\,kpc and a solar rotational speed $\Theta_\odot$ of $214\pm7$\,km\,s$^{-1}$ \citep{and2009}. 
Based on the preceding H{\sc i} absorption studies, \citet{and2009} resolved this ambiguity and favored the far distance.
However, as discussed by \citet{ort2013}, the spectroscopic observations of BD-14\,5014 provide better constraints that support the near distance for the O star, and we therefore adopt 3.8\,kpc in this study.
In this study we found that the full velocity width of the molecular gas associated with Sh2-48 is $\sim$5\,km\,s$^{-1}$, which provides a distance error of $\sim\pm$200\,pc.
The size of Sh2-48 is measured to be $\sim$11\,pc at the distance 3.8\,kpc.

To date, the spatial and velocity distributions of the molecular gas over the whole Sh2-48 have not been studied to date. 
\citet{and2009} analyzed the Boston University-Five College Radio Astronomy Observatory (BU-FCRAO) $^{13}$CO $J$=1--0 Galactic Ring Survey data (GRS; \cite{jac2006}), which were taken at a beam size of 46$''$, but did not focus on the gas distribution in Sh2-48. 
\citet{ort2013} observed a small part ($2'\times2'$) of Sh2-48 in the CO $J$=3--2 and CS $J$=7--6 emission lines to study a bright-rimmed cloud in Sh2-48.

In this study, we report on the molecular gas distribution in the CO $J$=1--0 emission lines for a large area including Sh2-48, which we obtained using the Nobeyama 45-m telescope at 21$''$ resolution, which corresponds to $\sim$0.39\,pc at 3.8\,kpc. 
This high-resolution CO dataset provides a wealth of information on the distribution and dynamics of the molecular gas in Sh2-48, which allows us to investigate high-mass star formation in this region.
The remainder of this paper is organized as follows.
In Section\,2 we describe the CO $J$=1--0 dataset used in the study. 
In Section\,3 we present the main results of our analyses of the dataset and comparisons with the results at other wavelengths. 
In Section\,4 we discuss our results and, finally, a summary is presented in Section\,5.
In the following discussion, the four points of the compass are defined based on the Galactic coordinates.

\section{Dataset}
We analyzed the $^{12}$CO, $^{13}$CO, and C$^{18}$O $J$=1--0 datasets obtained as a part of FOREST Unbiased Galactic plane Imaging survey using the Nobeyama 45-m telescope (FUGIN; see \cite{ume2017} for a full description of the observations and data reduction). 
FUGIN involved a large-scale Galactic plane survey using the FOur-beam REceiver System on the 45-m Telescope (FOREST; \cite{min2016}), a four-beam, dual-polarization, two-sideband receiver installed in the Nobeyama 45-m telescope. 
Typical system temperatures in FOREST were $\sim$250\,K for $^{12}$CO and $\sim$150\,K for $^{13}$CO and C$^{18}$O.
The backend system was the digital spectrometer ``SAM45'', which provided a bandwidth of 1\,GHz and a resolution of 244.14\,kHz. 
These figures correspond to 2,600\,km\,s$^{-1}$ and 1.3\,km\,s$^{-1}$, respectively, at 115\,GHz.  
The observations were made in the on-the-fly mode, and the output data were formatted into spatial and velocity grid-sizes of 8.5$''$ and 0.65\,km\,s$^{-1}$, respectively. 
Absolute intensity calibrations were performed by adopting Main beam efficiencies of $0.45\pm0.02$ and $0.43\pm0.02$ at 110 and 115\,GHz, respectively \citep{ume2017}.
To improve the sensitivity, we applied a two-dimensional Gaussian function to smooth the output data to a spatial resolution of 30$''$ for $^{12}$CO and $^{13}$CO and 45$''$ for C$^{18}$O. 
The final root-mean-square (r.m.s) noise fluctuations of the data were 0.5\,K, 0.3\,K, and 0.2\,K for $^{12}$CO, $^{13}$CO, and C$^{18}$O, respectively, at a channel resolution of 0.65\,km\,s$^{-1}$.

\begin{figure}
 \begin{center}
  \includegraphics[width=14cm]{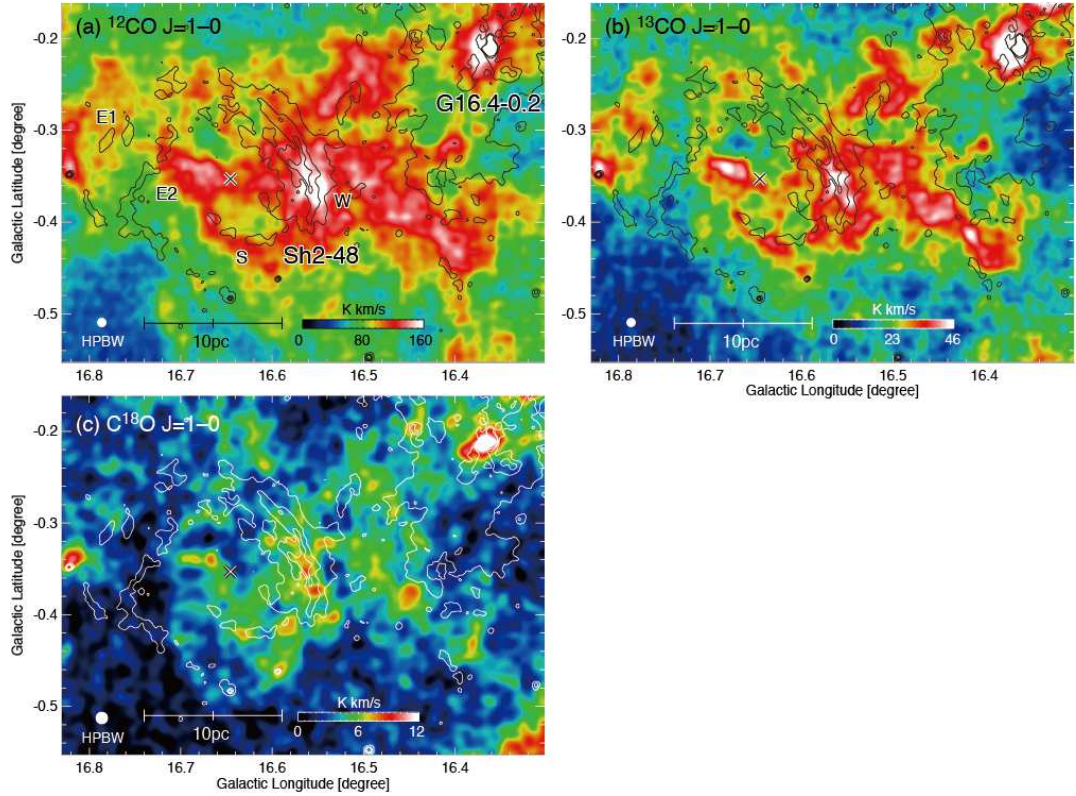}
 \end{center}
 \caption{(a) Intensity distributions of the CO emission integrated over the full velocity range of the Sh2-48 cloud, i.e., 38--53\,km\,s$^{-1}$. The three CO isotopologues, $^{12}$CO, $^{13}$CO, and C$^{18}$O, are presented in (a), (b), and (c), respectively. The cross indicates the position of BD-14\,5014, while the contours show the 8\,$\mu$m emission plotted at the same levels as in Figure\,\ref{fig:rgb}(a).}\label{fig:ii}
\end{figure}

\begin{figure}
 \begin{center}
  \includegraphics[width=17cm]{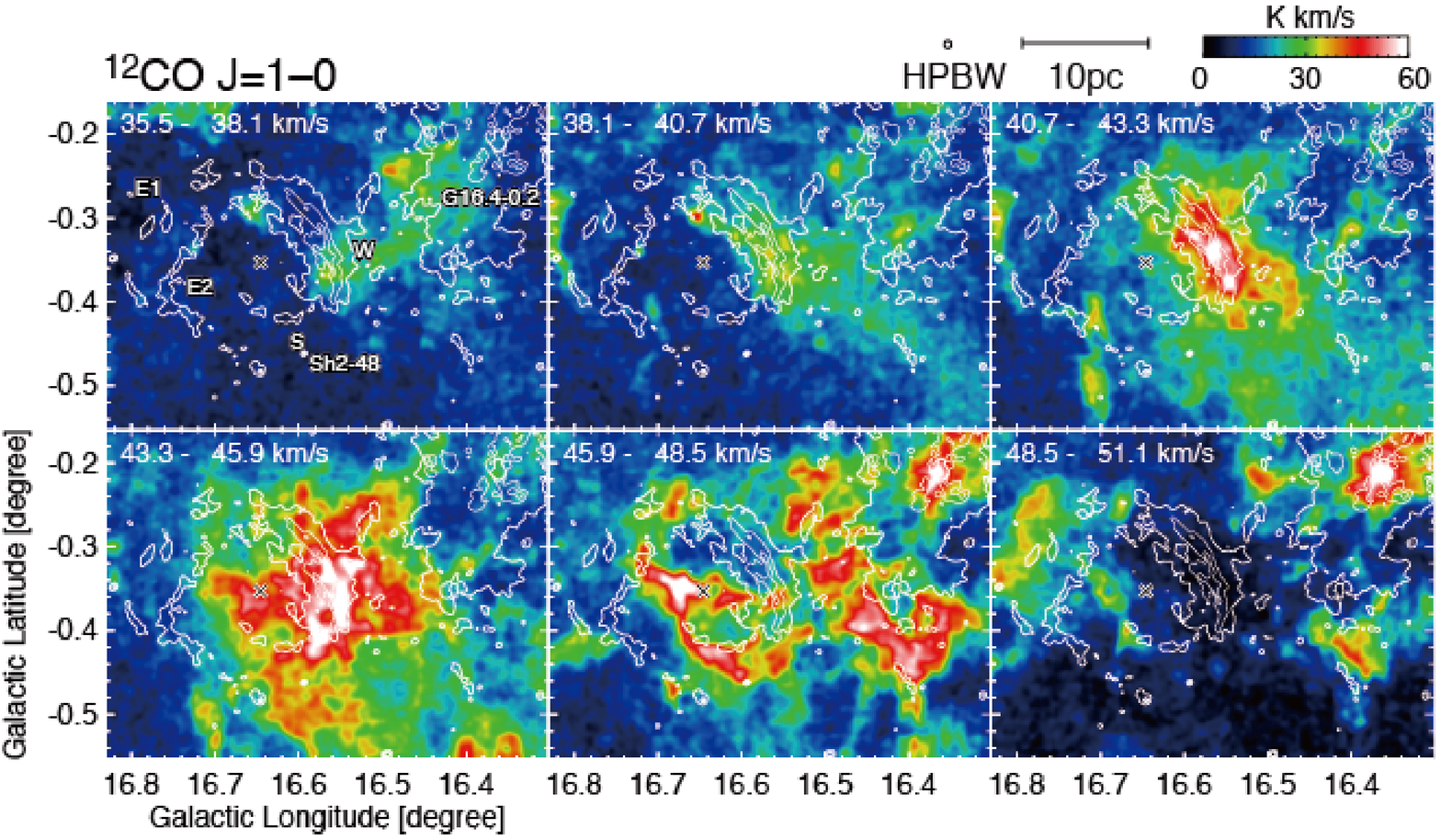}
 \end{center}
 \caption{Velocity channel map of the $^{12}$CO emission in Sh2-48. The cross indicates the position of BD-14\,5014, while the contours show the 8\,$\mu$m emission plotted at the same levels as in Figure\,\ref{fig:rgb}(a).}\label{fig:3x2_12}
\end{figure}

\begin{figure}
 \begin{center}
  \includegraphics[width=17cm]{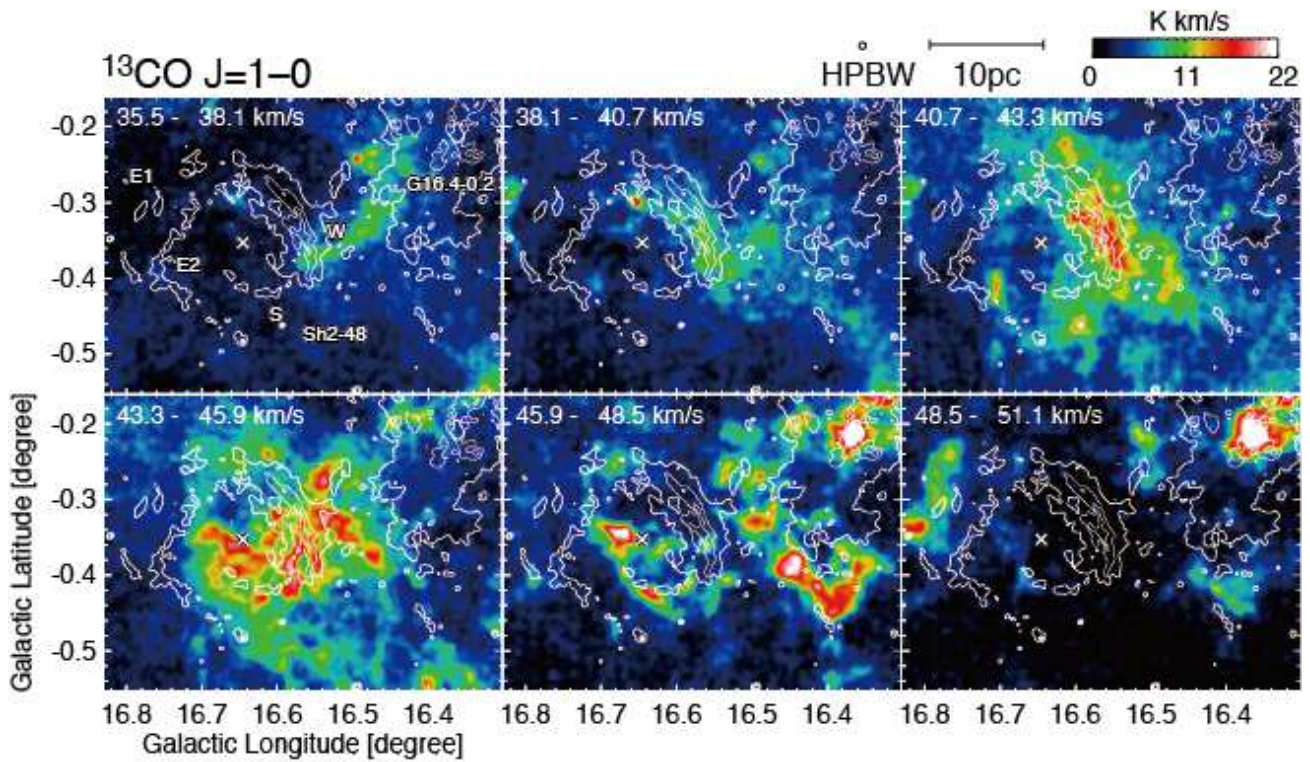}
 \end{center}
 \caption{Velocity channel map of the $^{13}$CO emission in Sh2-48. The cross indicates the position of BD-14\,5014, while the contours show the 8\,$\mu$m emission plotted at the same levels as in Figure\,\ref{fig:rgb}(a).}\label{fig:3x2_13}
\end{figure}

\section{Results}
\subsection{Spatial and velocity distribution of the molecular cloud in Sh2-48}

Figure\,\ref{fig:ii} shows the integrated CO emission maps over the 38--53\,km\,s$^{-1}$ velocity range in which the CO emission associated with Sh2-48 is prominent. 
The integrated intensities of the $^{12}$CO and $^{13}$CO emission lines indicate that the molecular gas associated with Sh2-48 extends for $\sim$20\,pc\,$\times$\,15\,pc centered on the bright CO emission at $(l,b)\sim(16\fdg56, -0\fdg35)$ and has filamentary structures with widths of $\sim$1--2\,pc and clumpy structures with sizes of $\sim$2--3\,pc.
The other bright CO emission seen to the northwest at $(l,b)\sim(16\fdg37, -0\fdg21)$ corresponds to an H{\sc ii} region complex other than Sh2-48 (hereafter G16.4-0.2). 
The H{\sc ii} regions included in G16.4-0.2 are compact in angular sizes, as is seen in Figure\,\ref{fig:rgb}(b). 
The C$^{18}$O emission in Sh2-48 is widely detected where the $^{12}$CO and $^{13}$CO emissions are bright, indicating the presence of dense gas with gas densities of $\sim10^3$--$10^4$\,cm$^{-3}$.

Unlike the 8\,$\mu$m filaments, the central bright part of the CO emission toward Sh2-48 spatially corresponds to filament W, while filament S is traced by the northern rim of a filamentary gas structure, as seen in the $^{12}$CO and $^{13}$CO maps in Figures\,\ref{fig:ii}(a) and (b), respectively.
It is not clear in the present figure even if the other two filaments, E1 and E2, have molecular gas counterparts.

Figures\,\ref{fig:3x2_12} and \ref{fig:3x2_13} show velocity channel maps of the $^{12}$CO and $^{13}$CO emissions with a velocity interval of $\sim$2.6\,km\,s$^{-1}$, respectively.
In the Appendix we include the velocity channel maps of the C$^{18}$O emission as supplementary material.
At around 40--43\,km\,s$^{-1}$, molecular gas is distributed in the direction corresponding to filament W, but is separated into two components to the east ($l\sim16\fdg6$--$16\fdg7$) and west ($l\sim16\fdg4$--$16\fdg5$) of filament W around 46--49\,km\,s$^{-1}$.
These lower and higher-velocity components are entangled at the intermediate velocities.
Molecular gas associated with G16.4-0.2 is prominent in a velocity range of $\sim$49--51\,km\,s$^{-1}$.
There is another CO component to the east of filament E1 located at $l\sim16\fdg8$ in the same velocity range.
As shown in Figure\,\ref{fig:pv}, this CO component appears to not be associated with Sh2-48.

\begin{figure}
 \begin{center}
  \includegraphics[width=17cm]{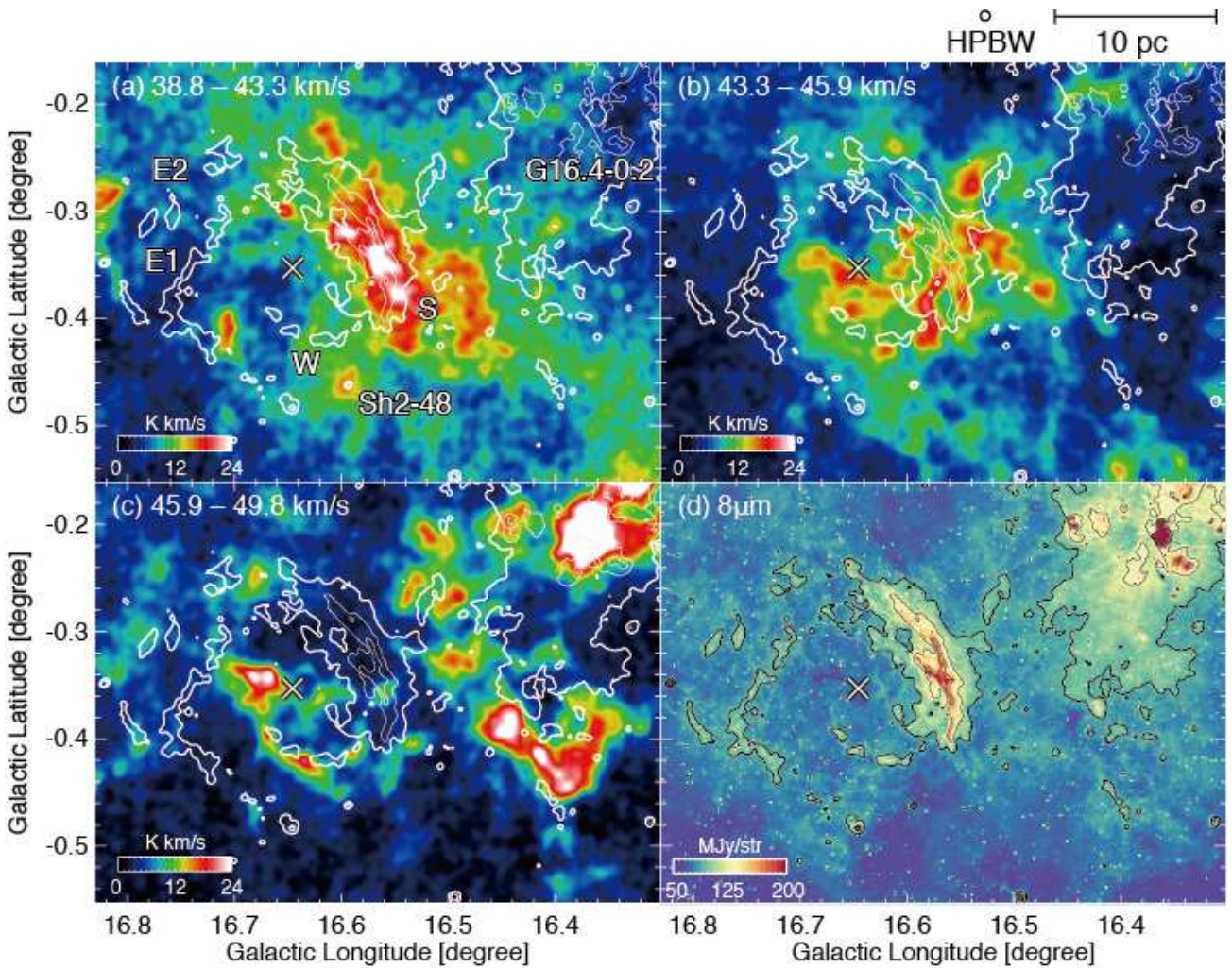}
 \end{center}
 \caption{(a)--(c) Integrated intensity distributions of the $^{13}$CO emission for the three velocity ranges of the Sh2-48 cloud. (d) {\it Spitzer}/GLIMPSE 8\,$\mu$m map presented as a contoured image. The cross indicates the position of BD-14\,5014, while the contours show the 8\,$\mu$m emission plotted at the same levels as in Figure\,\ref{fig:rgb}(a).}\label{fig:3vel}
\end{figure}

Figure\,\ref{fig:3vel}(a) shows the $^{13}$CO distributions of the lower-velocity component (hereafter, the LVC) for $\sim$39--43\,km\,s$^{-1}$. The two higher velocity components to the east and the west (hereafter, the HVC-E and HVC-W, respectively) are shown for $\sim$46--50\,km\,s$^{-1}$ in Figure\,\ref{fig:3vel}(c). 
The LVC has an elongated shape along the northeast-southwest direction, that traces filament W, while the HVC-E and HVC-W have fragmented gas components roughly aligned along the same direction as the LVC.
The molecular gas associated with filament S is included in the HVC-E.
These velocity components are merged in the intermediate velocity range in Figure\,\ref{fig:3vel}(b), resulting in an entangled filamentary distribution.

\begin{figure}
 \begin{center}
  \includegraphics[width=10cm]{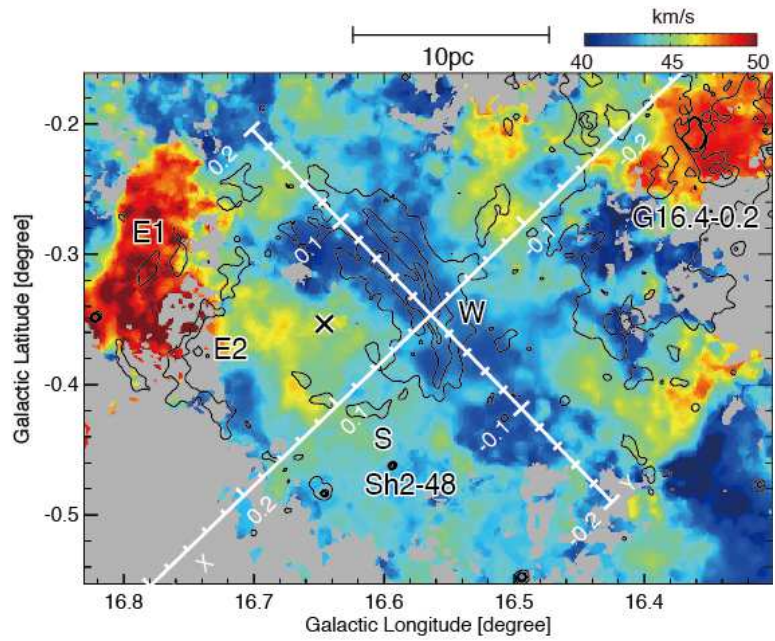}
 \end{center}
 \caption{First-moment map of the $^{13}$CO $J$=1--0 data. The $X$- and $Y$-axes and labels are used to define the position-velocity map coordinates in Figure\,\ref{fig:pv}.  The cross indicates the position of BD-14\,5014, while the contours show the 8\,$\mu$m emission plotted at the same levels as in Figure\,\ref{fig:rgb}(a).}\label{fig:mom}
\end{figure}

\begin{figure}
 \begin{center}
  \includegraphics[width=6.5cm]{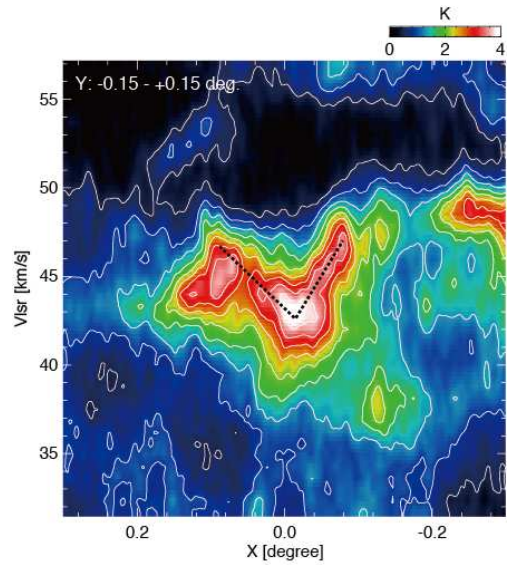}
 \end{center}
 \caption{Position-velocity diagram of $^{13}$CO emission along the $X$-axis of the coordinates defined in Figure\,\ref{fig:mom}. Contours start at 5$\sigma$ and repeat in intervals of 4$\sigma$, corresponding to 0.5\,K and 0.4\,K, respectively. }\label{fig:pv}
\end{figure}

Figure\,\ref{fig:mom} shows a first-moment map of the $^{13}$CO emission created for the velocity range 38--53\,km\,s$^{-1}$ using voxels with intensities higher than 5\,$\sigma$ ($\sim1.5$\,K).
The LVC is seen as low velocity area with clear boundary in this map, that is spatially coincident with filament W.
To investigate the velocity distribution of the molecular gas in more depth, we defined a new coordinate system based on the orientation of the LVC relative to Galactic coordinates.
We performed a linear fit to the $^{13}$CO integrated intensity map of the LVC shown in Figure\,\ref{fig:3vel}(a) using data points with integrated intensities larger than two-third of the peak intensity 27\,K\,km\,s$^{-1}$ and measured the angle of the best-fit line to the Galactic plane.
The derived angle, $\sim130^\circ$, was then used to define the $Y$-axis of the new coordinates as plotted in Figure\,\ref{fig:mom}.
The origin of the coordinates is set to the peak position of the $^{13}$CO integrated intensity map in Figure\,\ref{fig:3vel}(a) located at $(l,b,) = (16\fdg560, -0\fdg355)$.

Figure\,\ref{fig:pv} shows the position-velocity diagram of the $^{13}$CO emission along the $X$-axis of the new coordinate system, revealing velocity gradients in the molecular gas from the center ($X\sim0^\circ$) to both sides, following a ``V-shaped'' gas distribution (as guided by dashed lines).
The bottom of the ``V'' corresponds to the LVC, while the two tops of the ``V'' represent the HVC-E and HVC-S.
The velocity difference between the top and bottom of the V-shaped gas distribution is measured to be $\sim$5\,km\,s$^{-1}$, with the velocity gradients at the both sides of the V-shaped estimated at $\sim$5\,km\,s$^{-1}$ / 7\,pc\,$\sim$\,0.7\,km\,s$^{-1}$\,pc$^{-1}$.
The V-shaped gas distribution suggests that it is a single molecular cloud, comprising the LVC, HVCs, and intermediate CO features.
Hereafter, we refer to this molecular cloud as ``the Sh2-48 cloud''.
Note that the CO emission associated with G16.4-0.2 is connected to the Sh2-48 cloud, suggesting that they are located at a common distance. 
The CO component to the east of filament E1 is detected at 51--54\,km\,s$^{-1}$ at an $X$-coordinate of $\sim0\fdg1$--$0\fdg2$, and appears not to be connected to the Sh2-48 cloud.

\subsection{Physical parameters of the molecular cloud}
\begin{figure}
 \begin{center}
  \includegraphics[width=15cm]{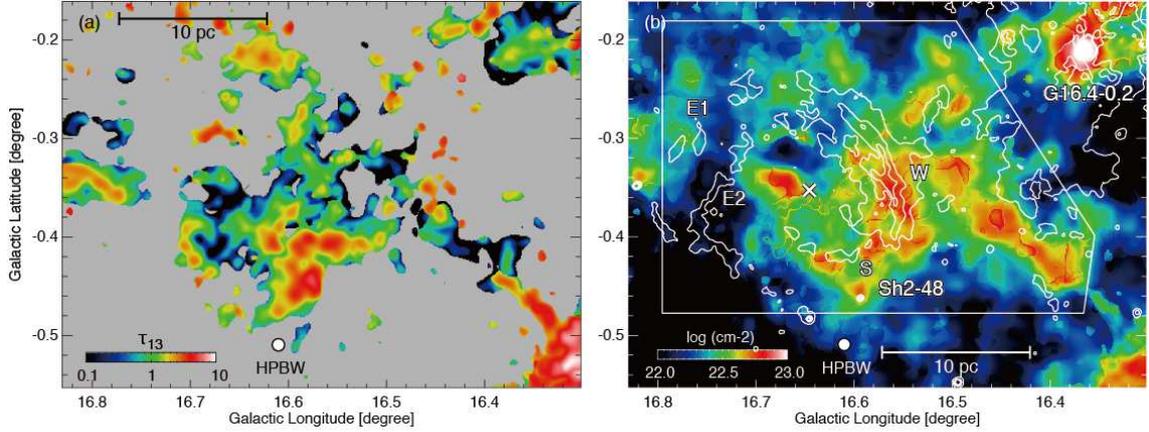}
 \end{center}
 \caption{(a) Peak $\tau_{13}$ distribution derived from the intensity ratio of $^{13}$CO to C$^{18}$O in the equation\,(\ref{eq:1}). (b) $N{\rm (H_2)}$ distribution computed using the equations (\ref{eq:3}) and (\ref{eq:4}). The cross indicates the position of BD-14\,5014, while the contours show the 8\,$\mu$m emission plotted at the same levels as in Figure\,\ref{fig:rgb}(a).}\label{fig:nh2}
\end{figure}

To estimate the molecular mass of the Sh2-48 cloud, we quantified its H$_2$ column density $N{\rm (H_2)}$ from the $^{13}$CO and C$^{18}$O data by assuming a common beam filling factor \citep{bou1997, won2008}.
First, we estimated the opacity of the two lines for each voxel ($x, y, v$) within the data cube corresponding to the velocity range 39--50\,km\,s$^{-1}$ (as determined from Figure\,\ref{fig:pv}) using the ratio of $^{13}$CO to C$^{18}$O main-beam brightness temperatures:
\begin{equation}
\frac{T(\rm{^{13}CO})}{T(\rm{C^{18}O})} = \frac{1-\exp (-\tau_{13})}{1-\exp (-\tau_{18})} = \frac{1-\exp (-\tau_{13})}{1-\exp (-\tau_{13}/6.2)}, \label{eq:1}
\end{equation}
where we assume a ratio of $^{13}$CO opacity $\tau_{13}$ to C$^{18}$O opacity $\tau_{18}$ of 6.2, based on the abundance ratio [$^{13}$C][$^{16}$O]/[$^{12}$C][$^{18}$O]\,=\,$6.2\pm0.6$ \citep{wil1994}.
To reduce the effects of noise, the spectra were binned to a velocity resolution of 1.95\,km\,s$^{-1}$.
The ratio $T$($^{13}$CO)/$T$(C$^{18}$O) was then computed for voxels with C$^{18}$O signal-to-noise ratios (S/N) higher than five.
In Figure\,\ref{fig:nh2}(a) we plot the peak $\tau_{13}$ distribution, which is seen to be typically $\tau_{13}$ of 0.5--3 in the Sh2-48 cloud. 
The southern part of the cloud around at $(l,b)\sim(16\fdg55$--$16\fdg62$, $-0\fdg48$--$-0\fdg40)$ show high values of $\tau_{13}$ of up to five.
We cannot find very high $\tau_{13}$ of 10 in the Sh2-48 cloud, at which it is difficult to accurately measure $\tau_{13}$ given noise levels of the CO spectra.


The excitation temperature $T_{\rm ex}$ was calculated using the $^{13}$CO data from the equation of radiative transfer:
\begin{equation}
T(^{13}{\rm CO}) \ = \ [J(T_{\rm ex}) - J(T_{\rm bg})] [1-\exp(-\tau_{13})], \label{eq:2}
\end{equation}
where $J(T) \equiv 5.29 / [\exp (5.29/T) - 1]$ and $T_{\rm bg} = 2.7$\,K.
Using the derived $T_{\rm ex}$ and $\tau_{13}$, the $^{13}$CO column densities $N{\rm (^{13}CO)}$ were calculated as follows \citep{gar1991}:
\begin{equation}
N{\rm (^{13}CO)}\ = \ 2.42 \times 10^{14} \frac{T_{\rm ex} + 0.88}{1- \exp(-5.29/T_{\rm ex})} \int \tau_{13} \, dv, \label{eq:3}
\end{equation}
where $v$ is measured in km\,s$^{-1}$.
Voxels with $\tau_{13}$ $<$ 0.1 were assigned $N{\rm (^{13}CO)}$ in the optically thin limit \citep{bou1997, won2008}:
\begin{equation}
N{\rm (^{13}CO)} \ = \ 2.42 \times 10^{14}  \frac{T_{\rm ex} + 0.88}{1- \exp(-5.29/T_{\rm ex})} \frac{1}{J(T_{\rm ex}) - J(T_{\rm bg})} \int T{\rm (^{13}CO)} \, dv, \label{eq:4}
\end{equation}
In this case the $T_{\rm ex}$ was calculated in each line-of-sight from the peak intensity of the $^{12}$CO spectrum with an assumption of optically thick emissions:
\begin{equation}
T_{\rm ex}\ = \ \frac{5.53}{\ln\{ 1 + 5.53 / (T(^{12}{\rm CO}) + 0.819) \} }, \label{eq:5}
\end{equation}
We also applied this procedure in the voxels with $T$($^{13}$CO)/$T$(C$^{18}$O) $>$ 6.2, at which we cannot measure $\tau_{13}$ from the equation \ref{eq:1}, or those at which only $^{13}$CO was detected.

Finally, the $N{\rm (H_2)}$ distribution of the Sh2-48 cloud was calculated by integrating the $N{\rm (^{13}CO)}$ data cube over a velocity range of 39--50\,km\,s$^{-1}$ assuming [H$_2$]/[$^{12}$CO] = $10^4$ \citep{fre1982, leu1984} and [$^{12}$C]/[$^{13}$C] = 53 \citep{wil1994}. Figure\,\ref{fig:nh2}(b) shows the derived $N{\rm (H_2)}$ distribution of the Sh2-48 cloud, which indicates that the cloud has a typical $N{\rm (H_2)}$ of $\sim4\times10^{22}$\,cm$^{-2}$, with several spots of size of 2--3\,pc having relatively high $N{\rm (H_2)}$ of $\sim6\times10^{22}$\,cm$^{-2}$.
The total molecular mass of the cloud can be estimated as $\sim3.8\times10^5$\,$M_\odot$ at 3.8\,kpc for the area enclosed by the white solid lines in Figure\,\ref{fig:nh2}(b).
The respective masses of the LVC and HVCs (HVC-E and HVC-W) were calculated toward the same area for the velocity ranges 39--43\,km\,s$^{-1}$ and 46--50\,km\,s$^{-1}$ as $\sim1.4\times10^5$\,$M_\odot$ and $\sim1.2\times10^5$\,$M_\odot$, respectively.
The mass included in the intermediate velocity range is $\sim1.2\times10^5$\,$M_\odot$.

As the $^{13}$CO and C$^{18}$O emissions show the extended spatial distributions compared to the telescope beam sizes (Figure\,\ref{fig:ii}), in the equation\,\ref{eq:1} we assume common beam filling factors between $^{13}$CO and C$^{18}$O.
If we tentatively assume 20\% smaller beam filling factors for C$^{18}$O than $^{13}$CO, the derived $\tau_{13}$ (and thus the $N{\rm (H_2)}$) decreases by $\sim$30--50\%.
The $\pm\sim200$\,pc uncertainties in the distance of Sh2-48 and $\pm\sim$5\% uncertainties in the Main beam efficiency of the telescope, which affects the $T_{\rm ex}$ estimates, will also change the inferred $N{\rm (H_2)}$ by $\sim\pm10\%$ in each.
Considering these effects, it can be esitmated that the total mass of the Sh2-48 cloud may be overestimated by up to 30\%, where the uncertainties in the $\tau_{13}$ account for $\sim$10\%, as the $\sim$20\% of the total cloud mass is accounted for by the $N{\rm (H_2)}$ computed using the equation\,\ref{eq:3} which includes $\tau_{13}$.

\subsection{Comparisons with other wavelengths}
We then compared the FUGIN CO dataset to images taken at other wavelengths.
As already indicated in Figures\,\ref{fig:3x2_12}--\ref{fig:3vel}, filaments W and S in the 8\,$\mu$m image show spatial correlations with the LVC and HVC-E, respectively.
As PAHs are radiatively excited at the surface of the molecular cloud, the brightness of filament S at the northern rim of the filamentary structure in the HVC-E indicates that the molecular gas is irradiated from the north, suggesting that the gas is located in the vicinity of BD-14\,5014.
On the other hand, although filament W shows no clear spatial offset with the LVC (which helps constrain the direction of the UV source), the strong 8\,$\mu$m emission in filament W suggest that it is excited by strong UV radiation from the massive star in Sh2-48.

\begin{figure}
 \begin{center}
  \includegraphics[width=15cm]{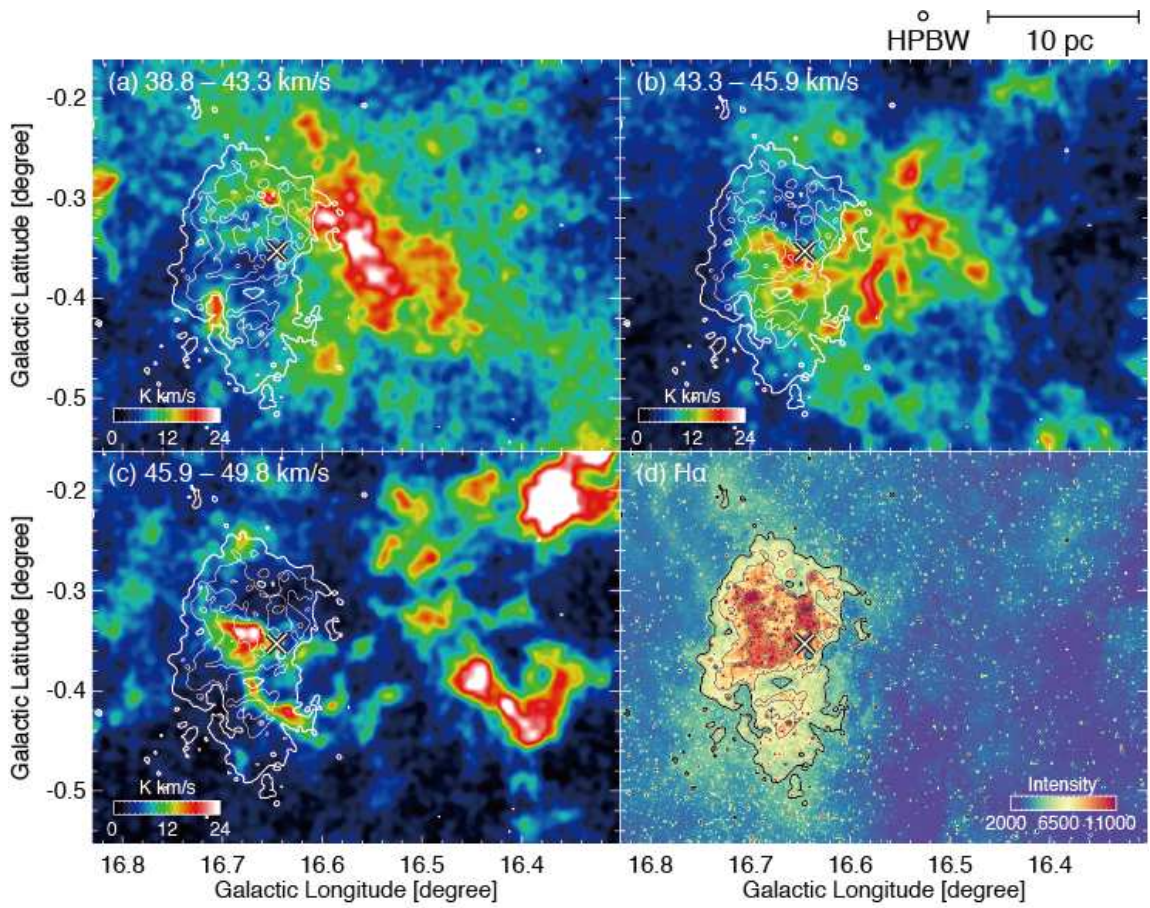}
 \end{center}
 \caption{(a)--(c) Comparisons of the $^{13}$CO integrated intensity maps with the SuperCOSMOS H$\alpha$ contour map for the three velocity ranges. (d) H$\alpha$ map presented as a contoured image \citep{hel2006}. Contours start at 5$\sigma$ and repeat in intervals of 2$\sigma$ of the local background noise level, which correspond to intensities of 5,000 and 2,000, respectively. Cross depicts BD-14\,5014.}\label{fig:comp1}
\end{figure}

Figure\,\ref{fig:comp1} shows comparisons between the $^{13}$CO data and the SuperCOSMOS H$\alpha$ map \citep{car2009} for the three velocity ranges.
In Figure\,\ref{fig:comp1}(a) the LVC demonstrates a distribution that is complementary to the H$\alpha$ emission, suggesting that the LVC is located in front of Sh2-48 along the line-of-sight, and that the H$\alpha$ emission is absorbed by dust grains in the LVC.
Two compact molecular components at $(l,b)\sim(16\fdg65, -0\fdg28)$ and $(16\fdg70, -0\fdg42)$ in the same velocity range coincide with intensity decreases in the H$\alpha$ map, with the former corresponding to the bright rimmed cloud studied by \citet{ort2013}, in which the authors discussed the interaction between the cloud and the H{\sc ii} region. 
On the other hand, in Figure\,\ref{fig:comp1}(c) we found no clear correlation between the HVC-E and H$\alpha$ emission, although the $N{\rm (H_2)}$ of the HVC-E is not much different from that in the LVC, suggesting that the HVC-E is located either at the inside or at toward the rear side of the H{\sc ii} region.
These signatures indicate that, even though they are connected in the position-velocity diagram (Figure\,\ref{fig:pv}), the LVC and HVC-E are located at the opposite sides of the H{\sc ii} region.

\begin{figure}
 \begin{center}
  \includegraphics[width=15cm]{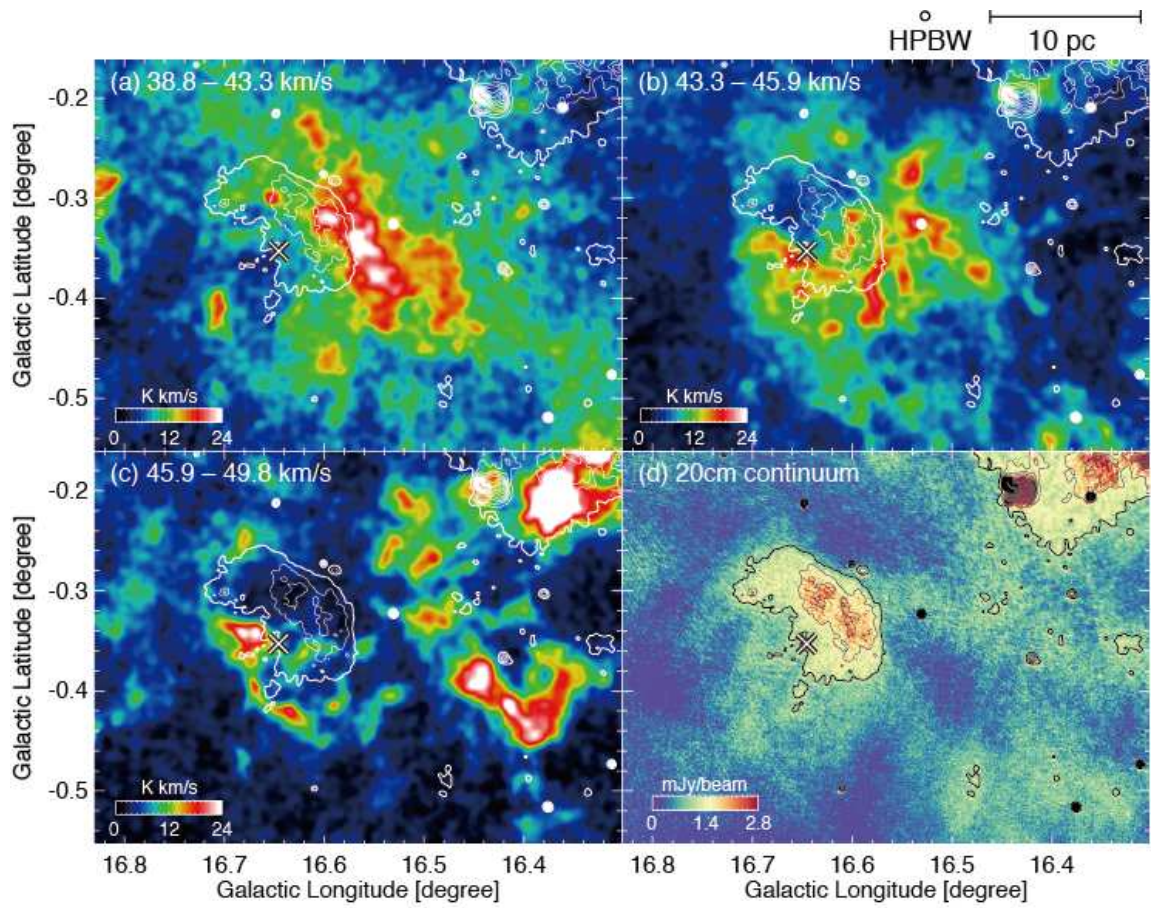}
 \end{center}
 \caption{(a)--(c) Comparisons of the $^{13}$CO integrated intensity maps with the MAGPIS 20\,cm data \citep{hel2006}. (d) 20-cm map presented as a contoured image. Contours start at 5$\sigma$ and repeats in intervals of 2$\sigma$ of the local background noise level, which correspond to 0.48 and 1.2\,mJy\,beam$^{-1}$, respectively. Cross depicts BD-14\,5014.}\label{fig:comp2}
\end{figure}

Comparisons between the $^{13}$CO data with the MAGPIS 20\,cm data are presented in Figure\,\ref{fig:comp2} \citep{hel2006}.
The 20\,cm emission is enhanced at the east side of the LVC in Figure\,\ref{fig:comp2}(a), while the gas in the intermediate velocity range in Figure\,\ref{fig:comp2}(b) appears to surround the southern rim of the 20\,cm distribution. 
The filamentary structure of the HVC-E in Figure\,\ref{fig:comp2}(c) is distributed to the east of the 20\,cm emission at $l\sim16\fdg6$--$16\fdg7$.  
These signatures indicate physical interaction of these velocity components with Sh2-48.

\begin{figure}
 \begin{center}
  \includegraphics[width=6cm]{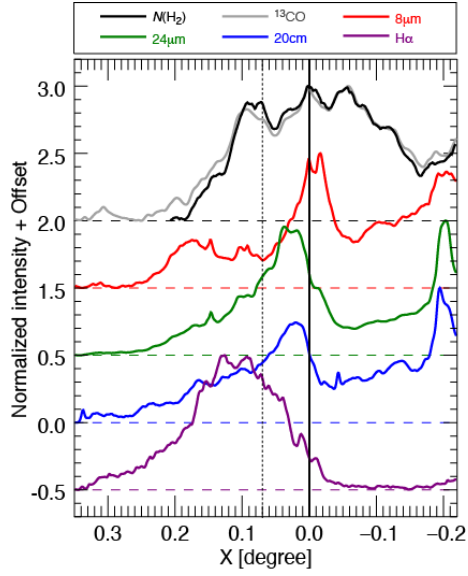}
 \end{center}
 \caption{Intensity distributions of the $N({\rm H}_2)$ (black), $^{13}$CO (gray), 8\,$\mu$m (red), 24\,$\mu$m (green), 20\,cm (blue), and H$\alpha$ (purple) emission along the $X$-axis of the coordinates defined in Figure\,\ref{fig:mom}, with the data averaged over a $Y$-range of $-0\fdg1$--$+0\fdg1$. The minimum and maximum values of each intensity distribution are normalized to zero and one, respectively, over the presented $X$-range. The solid line indicates $X$-coordinate\,=\,0\fdg0, which corresponds to the peak of the $^{13}$CO emission, while the dotted line indicates the position of BD-14\,5014.}\label{fig:rotplot}
\end{figure}

Figure\,\ref{fig:rotplot} shows the intensity distributions of the $N({\rm H}_2)$, $^{13}$CO, 8\,$\mu$m, 24\,$\mu$m, 20\,cm, and H$\alpha$ emission along the $X$-axis of the coordinates, which is defined in Figure\,\ref{fig:mom}.
The 8\,$\mu$m peak corresponds closely with the peak of the $N({\rm H}_2)$ in the LVC.
The 24\,$\mu$m profile, which probes warm dust grains, is similar to the 20\,cm profile, and both peaks are shifted toward the positive direction of the 8\,$\mu$m peak by $\sim$0\fdg03 in $X$-coordinate, which corresponds to $\sim$1.8\,pc at 3.8\,kpc.
The significant difference between H$\alpha$ and 20\,cm profiles can be interpreted as indicating that the H$\alpha$ emission around the 20\,cm peak is absorbed by the dust grains located in front of the H{\sc ii} region along the line-of-sight, as indicated in Figure\,\ref{fig:comp1}. 

\begin{figure}
 \begin{center}
  \includegraphics[width=8cm]{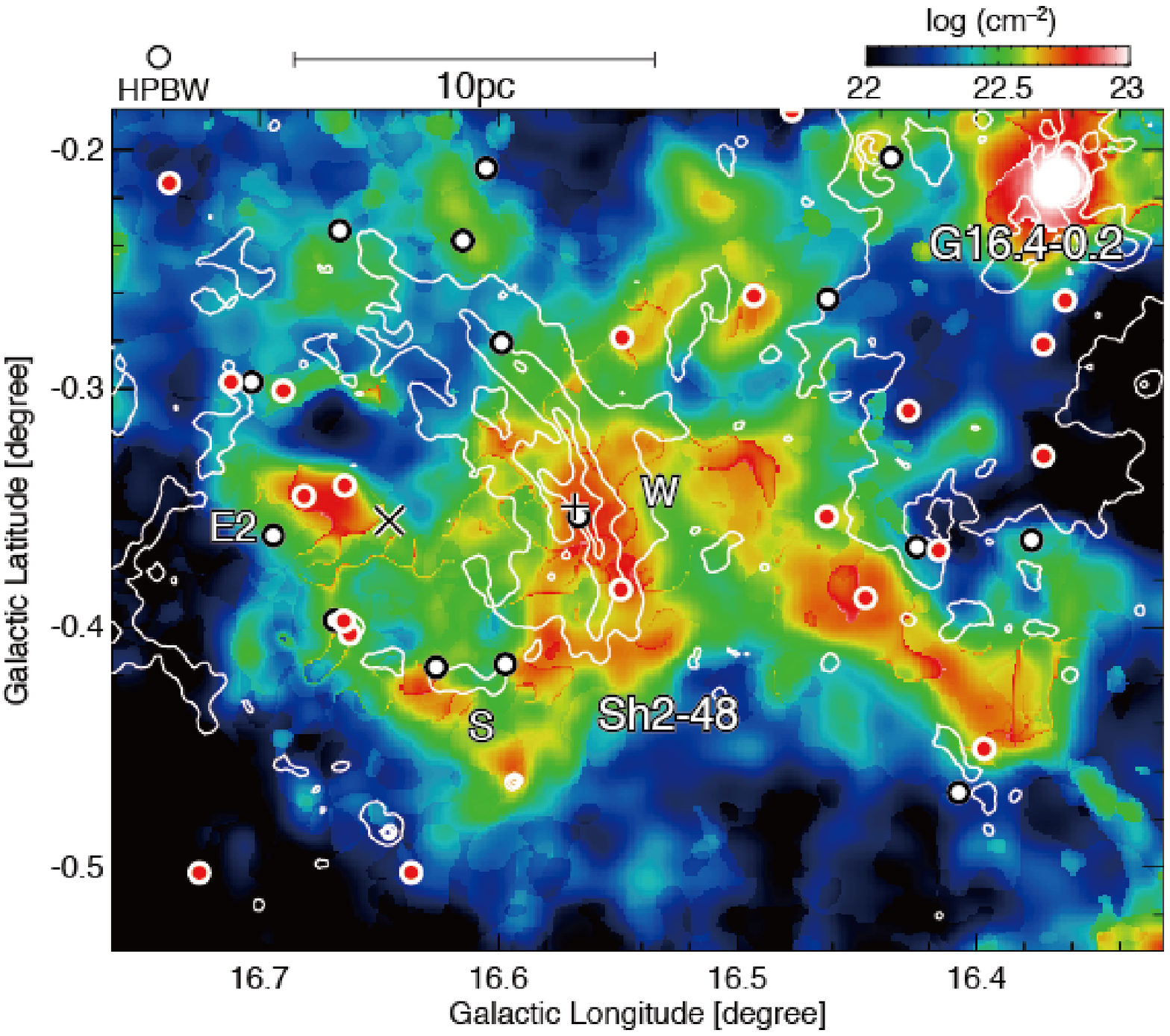}
 \end{center}
 \caption{Distributions of red sources identified using {\it Spitzer} (red circles, \cite{rob2008}) and the YSO candidates from the {\it AKARI} observations (white circles, \cite{tot2014}) superimposed on the $N({\rm H}_2)$ distribution in Figure\,\ref{fig:nh2}(b). The YSOs identified in a $20'\times20'$ box centered on the cloud center, $(l,b)\sim(16\fdg56, -0\fdg35)$, are plotted. White contours show the 8\,$\mu$m emission plotted at the same levels as in Figure\,\ref{fig:rgb}(a). Large cross and small cross indicate the positions of BD-14\,5014 and possible candidate source of the additional O star (see text), respectively.}\label{fig:yso}
\end{figure}

Figure\,\ref{fig:yso} superimposed the young stellar objects (YSO) candidates identified using {\it Spitzer} (red circles, \cite{rob2008}) and {\it AKARI} (white circle, \cite{tot2014}) observations onto the $N{\rm (H_2)}$ map shown in Figure\,\ref{fig:nh2}(b). 
\citet{rob2008} indicated that their {\it Spitzer}-derived catalog comprises 50--70\% YSOs and 30--50\% asymptotic giant branch stars. 
Figure\,\ref{fig:yso} shows that there are several YSO candidates coinciding with the clumpy structures of higher $N({\rm H}_2)$ of $\sim6\times10^{22}$\,cm$^{-2}$ embedded in the Sh2-48 cloud, suggesting that, if these are YSOs, the Sh2-48 cloud harbors on-going star formation, however, in the present observations we found no signs of massive YSO or protostars.

\section{Discussion}
In the previous section, we analyzed the spatial and velocity distributions of the Sh2-48 cloud and showed that the cloud has a V-shaped gas distribution in its position-velocity diagram (Figure\,\ref{fig:pv}).
In the lower-velocity range of around 42\,km\,s$^{-1}$, which corresponds to the bottom of the V-shaped gas distribution, the LVC of the Sh2-48 cloud, which is elongated nearly perpendicular to the Galactic plane, is spatially coincident with the bright filament W in the 8\,$\mu$m image.
The absorption in the H$\alpha$ image indicates that the LVC is located in front of Sh2-48.
In the higher-velocity range of around 47\,km\,s$^{-1}$, the molecular gas of the cloud is separated into the eastern and western sides of the LVC, or the HVC-E and HVC-W, which correspond to the two tops of the V-shaped distribution.
The HVC-E shows spatial correlation with the 8\,$\mu$m and 20\,cm data, indicating physical interaction with the H{\sc ii} region, but shows no clear spatial correlation with the H$\alpha$ data, suggesting that it is located toward the inside or rear side of the H{\sc ii} region.

In this section, we discuss the origin of the velocity gradients of gas in the Sh2-48 cloud and how it is related to the formation of the O star in Sh2-48.

\subsection{Formation timescale of the H{\sc ii} region in Sh2-48}
\begin{figure}
 \begin{center}
  \includegraphics[width=11cm]{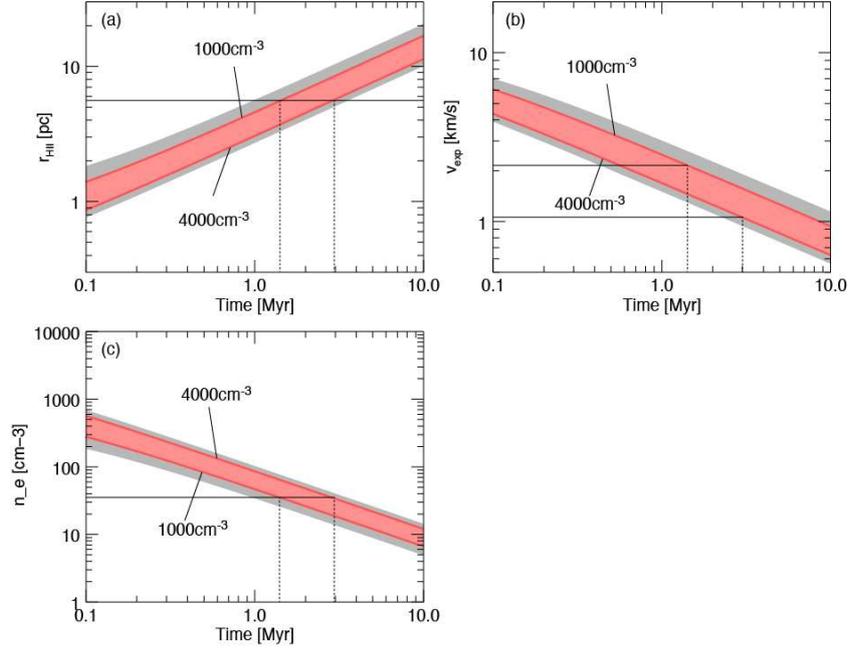}
 \end{center}
 \caption{Evolutionary tracks of expanding H{\sc ii} regions based on the Spitzer D-type expansion equation. The curves filled in red indicate the tracks of (a) $r_{\rm HII}$, (b) $v_{\rm exp}$, and (c) $n_{\rm e}$ for $n_0$ of $1000$--$4000$\,cm$^{-3}$. Gray areas indicate the $\pm50$\% errors of the individual tracks. The horizontal solid line in (a) indicates the observed size of Sh2-48. The vertical dashed lines depict the timescales corresponding to the observed size of Sh2-48. The tracks in (a)--(c) are calculated using the equation (9), the derivative of the equation (9), and the equation (6) of \citet{bis2015}, respectively. }\label{fig:hii}
\end{figure}

We analyzed the formation timescale of the H{\sc ii} region in Sh2-48 by measuring the flux of the Lyman continuum photons $N_{\rm Ly}$ using the 20\,cm image in Figure\,\ref{fig:rgb}(b).
Assuming the electron temperature of 8000\,K, we integrated the flux over all pixels having more than 1\,mJy\,beam$^{-1}$ to derive $N_{\rm Ly} \sim10^{48.92}$\,photons\,s$^{-1}$ \citep{kur1994}.
We also measured the typical emission measure and $n_{\rm e}$ of Sh2-48 as $\sim1.5\times10^4$\,pc\,cm$^{-6}$ and $\sim$40\,cm$^{-3}$, respectively, using the 20\,cm data; these figures are nearly consistent with the estimate by \citet{ort2013}. 
The estimated flux was then used to derive the evolution of the H{\sc ii} region based on an H{\sc ii} region model.
Figure\,\ref{fig:hii} plots evolutionary tracks of H{\sc ii} regions based on the analytical model of D-type expansion developed by \citet{spi1978} (see also \cite{bis2015} for more information on this simple H{\sc ii} region expansion model). 
We here assume $N_{\rm Ly} = 10^{48.92}$\,photons\,s$^{-1}$, $T_{\rm e} = 8000$\,K, and a speed of sound $c_{\rm i} = 10$\,km\,s$^{-1}$. 
In Figures\,\ref{fig:hii}(a), (b), and (c) the filled red areas indicate the tracks of the radius $r_{\rm HII}$, expansion velocity $v_{\rm exp}$, and the electron density $n_{\rm e}$ of the H{\sc ii} region for initial gas densities $n_0$ of $1000$--$4000$\,cm$^{-3}$, respectively.
The $n_0$ are measured assuming the sizes of the LVC and HVCs of $\sim$3--10\,pc and the typical $N{\rm (H_2)}$ of the Sh2-48 cloud of $\sim4\times10^{22}$\,cm$^{-2}$, as discussed in Section 3.2. 
The gray areas in the three panels indicate the $\pm50$\% errors in the tracks caused by uncertainties in $n_0$ and $N_{\rm Ly}$.
The horizontal solid line in Figure\,\ref{fig:hii}(a) indicates the observed $r_{\rm HII}$ of Sh2-48, $\sim$5.5\,pc, which corresponds to the evolutionary timescale of Sh2-48 ranging from $\sim$1.3--3\,Myr.
The inferred timescale indicates the $v_{\rm exp}$ and $n_{\rm e}$ of $\sim1$--$2$\,km\,s$^{-1}$ and $\sim$35\,cm$^{-3}$ in Figures\,\ref{fig:hii}(b) and (c), respectively.
The $v_{\rm exp}$ is smaller than the full velocity width of the Sh2-48 cloud of $\sim$5\,km\,s$^{-1}$, while the $n_{\rm e}$ is consistent with our measurement from the 20\,cm data.

\subsection{Expansion of the H{\sc ii} region}
In the following subsections, we discuss the components and origins of the observed velocity gradients of the Sh2-48 cloud  shown in Figure\,\ref{fig:pv}.
A reasonable approach to explaining these velocity gradient is expansion driven by the over-pressured H{\sc ii} region.
As discussed in the previous subsection, in this case the ambient gas would have an expansion velocity of $\sim$1--2\,km\,s$^{-1}$.
If we assume a spherical expansion of the H{\sc ii} region, and thus the ambient gas, a velocity gradient would appear between the exciting source and the rim of the H{\sc ii} region.
However, as shown in Figures\,\ref{fig:mom} and \ref{fig:pv}, the velocity gradients in the Sh2-48 cloud are observed even in the HVC-W, which is located at the outside the H{\sc ii} region, suggesting that the observed V-shaped gas distribution in the position-velocity diagram cannot be explained sorely by expansion of the H{\sc ii} region. 
Although G16.4-0.2 would be a likely candidate H{\sc ii} region for driving the HVC-W distribution, we found no evidence of physical interaction between these two.
The H{\sc ii} regions embedded within G16.4-0.2 are as small as $\sim$2--3\,pc in the 20\,cm image, and show no spatial correlations with the HVC-W (see Figures\,\ref{fig:rgb}(b) and \ref{fig:comp2}).
As shown in Figure\,\ref{fig:3vel}(c), the eastern boundary of the diffuse 8\,$\mu$m emission of G16.4-0.2 traces the western rim of the HVC-W, suggesting a potential physical interaction.
However, as the 8\,$\mu$m emission is not enhanced at the western rim of the HVC-W (Figure\,\ref{fig:3vel}(d)), this configuration is more likely attributable to the fact that the HVC-W is located in front of the diffuse 8\,$\mu$m emission, than it is to UV radiation from G16.4-0.2 illuminating its surface. 

\begin{figure}
 \begin{center}
  \includegraphics[width=10cm]{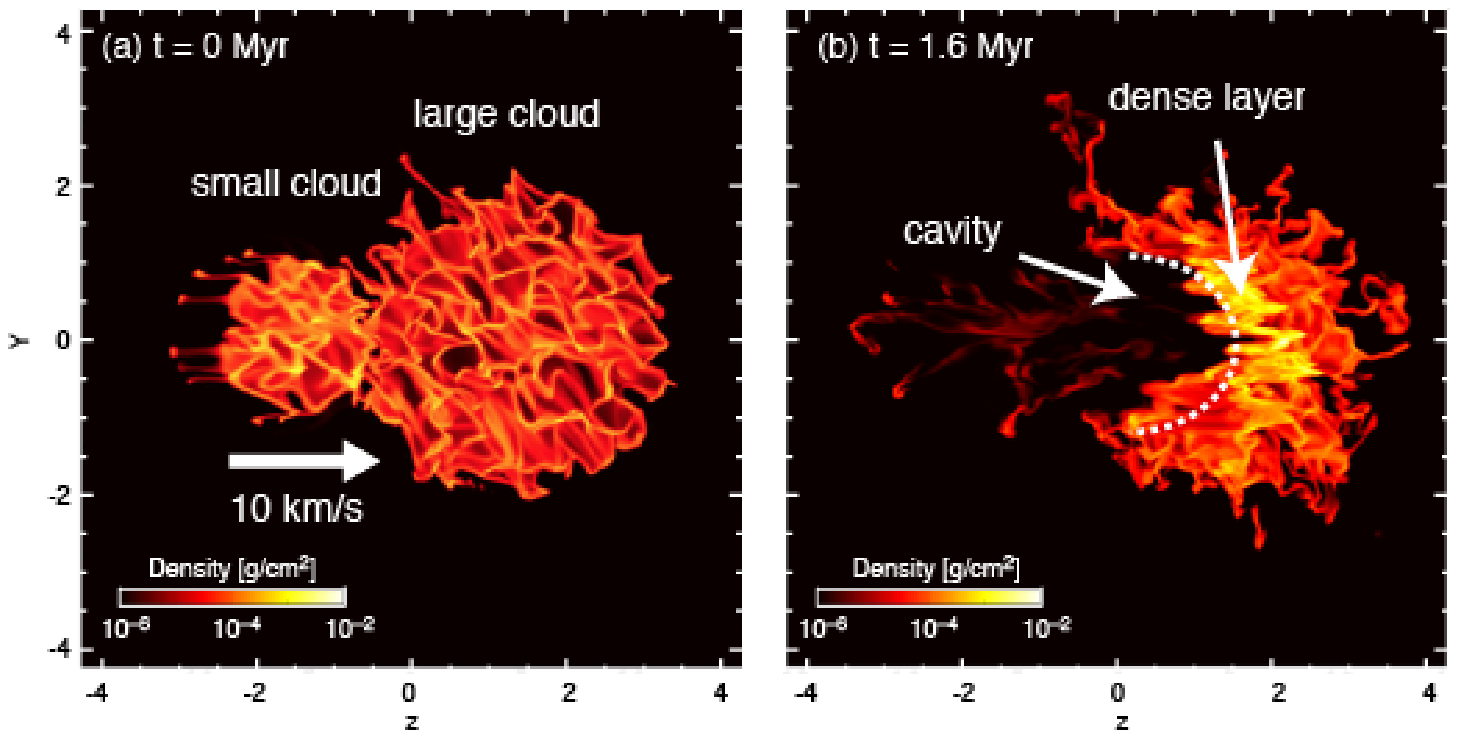}
 \end{center}
 \caption{Surface density plots of the 10\,km\,s$^{-1}$ CCC simulations by \citet{tak2014}. The clouds prior to collision are shown in (a), and a snapshot of the collision, in which the time corresponds to the maximum number of formed cores, is shown in (b) (see \cite{tak2014} for details). Both two snapshots are presented at a viewing angle perpendicular to the collision axis. The labels on the x, y, and z axes are normalized to the radius of the small cloud, 3.5\,pc. The maps shown in (a) and (b) are recomposed from the left panel of Figure\,7 and the rihgt-bottom panel of Figure\,6 in \citet{tak2014}, respectively.}\label{fig:ccc1}
\end{figure}

\begin{figure}
 \begin{center}
  \includegraphics[width=10cm]{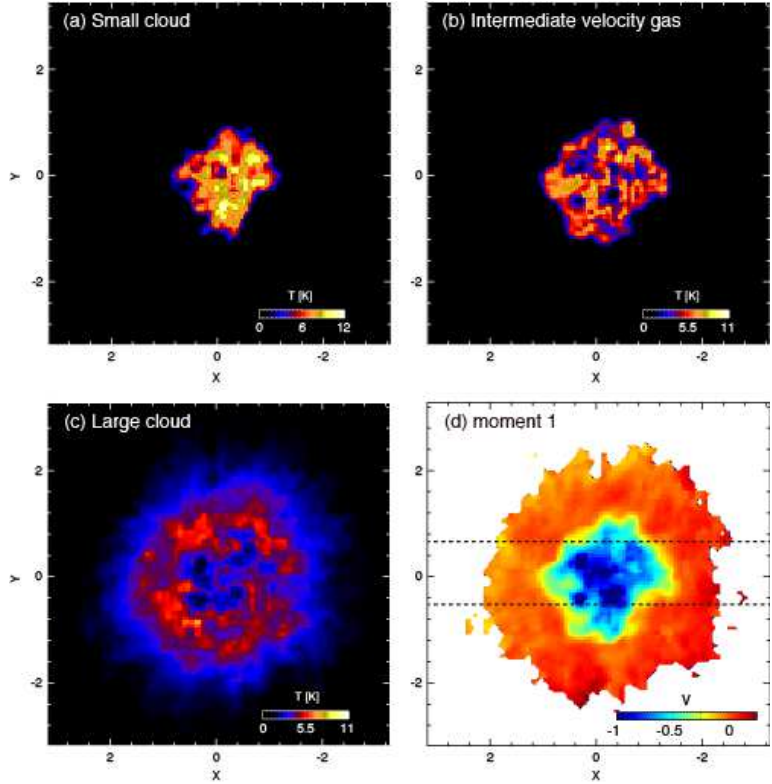}
 \end{center}
 \caption{(a)--(c) Projections of spatial distributions of the two colliding clouds onto the sky based on the CCC model data shown in Figure\,\ref{fig:ccc1}(b) using the synthetic CO $J$=1--0 observational data calculated by \citet{haw2015} are presented for three velocity ranges, which are indicated as dashed lines in Figure\,\ref{fig:ccc3}. The small and large clouds are shown in (a) and (c), respectively, while the CO emission at the intermediate velocity between the two clouds are shown in (b). The viewing angle of the synthetic observations is set to parallel to the collision axis. (d) First-moment map of the synthetic CO $J$=1--0 data. The color indicates the radial velocity of the gas normalized by the velocity separation between the large and small clouds of 5\,km\,s$^{-1}$. Dashed lines indicate the integration range in Y-axis in Figure\,\ref{fig:ccc3}.}\label{fig:ccc2}
\end{figure}

\subsection{Cloud-cloud collision (CCC) model}
Here, we propose a CCC model as an alternative idea to explain the observed V-shaped velocity distribution in the position-velocity diagram.
Figure\,\ref{fig:ccc1} shows surface density plots of CCC simulations by \citet{tak2014}, in which a collision between two dissimilar clouds at 10\,km\,s$^{-1}$ is presented. 
During the collision, the small cloud forms a cavity in the large cloud, forming a densely compressed, turbulent layer at the interface of the collision as shown in Figure\,\ref{fig:ccc1}(b).
Figure\,\ref{fig:ccc2} shows projections of spatial distributions of the two colliding clouds onto the sky based on the CCC model data shown in Figure\,\ref{fig:ccc1}(b) using the synthetic CO $J$=1--0 observational data calculated by \citet{haw2015}.
The line-of-sight of the synthetic observations is set to parallel to the collision axis.
The large cloud seen in Figure\,\ref{fig:ccc2}(a) has a ring-like morphology, with an inner radius corresponding to the radius of the small cloud in Figure\,\ref{fig:ccc2}(c), indicating the clouds' complementary distribution.
At the intermediate-velocity range shown in Figure\,\ref{fig:ccc2}(b), turbulent gas excitation is seen at the collisional interface, connecting the colliding clouds in the position-velocity diagram, and forming a bridge feature. 
Figure\,\ref{fig:ccc2}(d) shows the first-moment map of the synthetic CO data, demonstrating how the combination of the complementary distribution and bridge feature create a velocity gradient between the inside and outside of the collision region.

\citet{tor2017} observed the CO $J$=1--0 and 3--2 emission lines in the Galactic H{\sc ii} region M20, and based on comparisons between their observations and CCC numerical simulations by \citet{tak2014} and \citet{haw2015b}, discussed that ``spatially complementary distribution between two colliding clouds'' and ``bridge feature which connect the two colliding clouds in the position-velocity diagram' are  signatures of the molecular gas characteristic of young CCC regions.

\subsection{CCC and O star formation in Sh2-48}
\begin{figure}
 \begin{center}
  \includegraphics[width=16cm]{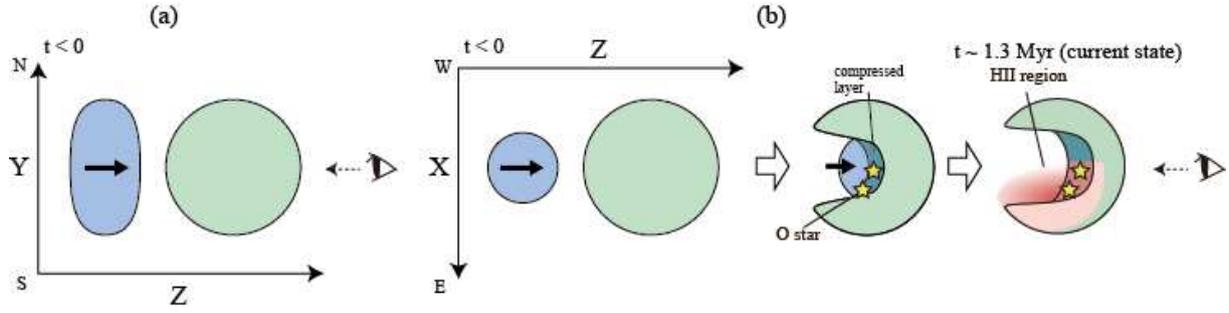}
 \end{center}
 \caption{Schematic of the CCC in Sh2-48, that triggered formation of the O stars. (a) The two clouds prior to the collision on the z-y plane. (b) The two clouds on the z-x plane. Time evolution of the clouds is shown from left to right. }\label{fig:ccc4}
\end{figure}

\begin{figure}
 \begin{center}
  \includegraphics[width=7cm]{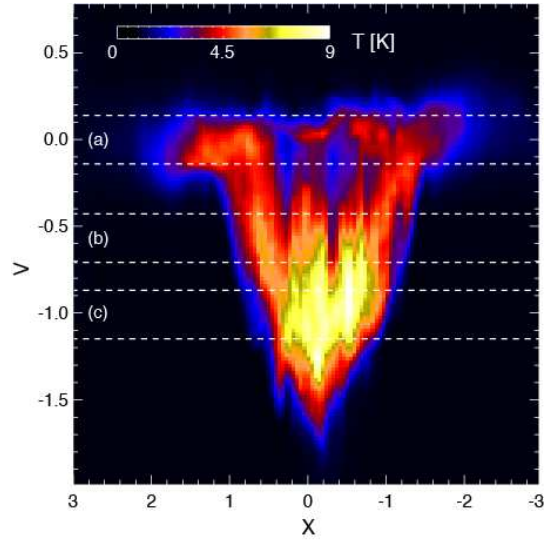}
 \end{center}
 \caption{Position-velocity diagram of the synthetic CO $J$=1--0 data of the collision model shown in Figure\,\ref{fig:ccc2}. The integration range in Y-axis is indicated as dashed lines in Figure\,\ref{fig:ccc2}(d). The labels of the X- and V-axes are normalized by the radius of the small cloud (3.5\,pc) and the velocity difference between the small cloud and the large cloud  (5\,km\,s$^{-1}$), respectively. The velocity ranges used for Figures\,\ref{fig:ccc2}(a)--(c) are indicated as dashed lines. }\label{fig:ccc3}
\end{figure}

Following the analysis in Section\,4.3 above, the observed V-shaped gas distribution in the position-velocity diagram of the Sh2-48 cloud (Figure\,\ref{fig:pv}) can be interpreted as the result of a collision between two clouds.
Figure\,\ref{fig:ccc4} shows our CCC scenario for Sh2-48 schematically. 
Whereas the case shown in Figures\,\ref{fig:ccc1}(b) and \ref{fig:ccc2} assumes the collision of two spherical clouds, we propose a collision between a vertically elongated, cylindrical cloud and a spherical cloud, as illustrated from the point of view of an observer parallel to the collision axis. 
In this case, the cylindrical cloud corresponds to the LVC.
In the proposed collision, the spherical cloud would have been separated into two components by the cylindrical cloud, and the observed HVC-E and HVC-W can be interpreted as these components.
Figure\,\ref{fig:ccc3} shows a position-velocity diagram of the synthetic CO $J$=1--0 data, with the $Y$-range of the diagram set to just cover the height of the small cloud, as indicated as dashed lines in Figure\,\ref{fig:ccc2}(d).
The CO emission in Figure\,\ref{fig:ccc3} shows a ``V-shaped'' velocity distribution, resembling the CO distribution in the observations shown in Figure\,\ref{fig:pv}.

The intermediate-velocity gas in the V-shaped gas distribution, which is attributed to turbulent gas excitation by the collision under the CCC scenario, is observable as long as the collision continues.
The expanded distribution of the C$^{18}$O emission in the Sh2-48 cloud shown in Figure\,\ref{fig:ii}(c) is consistent with this assumption, and suggests strong compression of gas by the collision.
The synthetic CO data show that the colliding regions of the gas comprise the network of filaments shown in Figure\,\ref{fig:ccc2}(b), which is consistent with the observed dense gas filamentary structures in the Sh2-48 cloud. 
The magnetohydrodynamical simulations of CCC by \citet{ino2017} also resulted in the formation of filamentary structures in the regions of collision.

In Section\,4.2, the timescale of Sh2-48 formation was measured as $\sim$1.3--3\,Myr for model cases with $n_0$ of $1000$--$4000$\,cm$^{-3}$.
The currently observed collision velocity in Sh2-48 is be measured to be $\sim$5\,km\,s$^{-1}$ from the velocity difference between the top and bottom of the V-shaped velocity distribution (Figure\,\ref{fig:pv}); as the velocity would have decreased during the collision, the initial collisional velocity would have been much faster.
If we tentatively assume a time-average collision velocity in Sh2-48 of 6\,km\,s$^{-1}$, the travel distance of the collision can be calculated to be $\sim$7.8--18\,pc for the timescale of $\sim$1.3--3\,Myr.
As the travel distance of 18\,pc is much larger than the size of the Sh2-48 cloud, in which case the collision would no longer continues. 
By contrast, the travel distance of 7.8\,pc is smaller than the projected size of the Sh2-48 cloud, and the formation timescale of Sh2-48 of $\sim1.3$\,Myr is probably consistent with the present observations suggesting that the CCC which triggered formation of the high-mass stars in Sh2-48 is still on going.
Further star formation, including the formation of massive stars, can possibly be triggered in Sh2-48 in the future, although, as mentioned in Section\,3.3, the Sh2-48 cloud currently shows no signs of massive YSOs or massive protostars.
Observations with high spatial and velocity resolutions will allow us to investigate the detailed distributions and dynamics of the gas in the filaments and clumps in the Sh2-48 cloud.

Here, we should discuss the possibility of hidden high-mass star(s) in Sh2-48.
As mentioned in Section\,1, only one O star---BD-14\,5014---has been identified in Sh2-48.
However, the $N_{\rm Ly}$ measured in Section\,4.1 corresponds to a single O6.5 star or four O9.5 stars \citep{pan1973}.
A single O9.5V star should only account for a fraction of the derived $N_{\rm Ly}$, suggesting that there are one or more additional O stars in Sh2-48, that are much brighter than BD-14\,5014.
A possible site of additional O stars is the vicinity of the 20\,cm, 8\,$\mu$m, and 24\,$\mu$m peaks, which is also close to the peak of the $^{13}$CO emission, as shown in Figure\,\ref{fig:rotplot}.
The figure shows no significant enhancements in these emissions in the direction of BD-14\,5014; instead, they are brightest to the west of BD-14\,5014 by 2--4\,pc in $X$-coordinate, where $N{\rm (H_2)}$ is as large as $\sim6\times10^{22}$\,cm$^{-2}$ around their respective peaks, corresponding to a visual extinction $A_{\rm V}$ of $\sim$32\,mag, assuming the relations $N({\rm H+H_2}) = 5.8\times10^{21} E(B-V)$\,cm$^{-2}$\,mag$^{-1}$ and $A_{\rm v} = 3.1 E(B-V)$ \citep{boh1978}. 
This high $A_{\rm V}$ might veil a bright O star.

In the Two Micron All Sky Survey (2MASS) point source catalog \citep{cut2003}, BD-14\,5014 has a $K_{\rm S}$ of $\sim$8.3\,mag, with $J - H$ and $H-K_{\rm S}$ colors of $\sim$0.2\,mag and $\sim$0.06\,mag, respectively. The 2MASS catalog also indicates several tens of 2MASS sources having brighter $K_{\rm S}$ in the direction of Sh2-48. 
Among them, a possible candidate source of the additional O star in Sh2-48 is found at $(l,b)\sim(16\fdg568, -0\fdg347)$, which corresponds to the brightest spot of the 8\,$\mu$m emission in filament W (Figure\,\ref{fig:yso}). 
The source, 2MASS J18221198-1441049, has a $K_{\rm S}$, $J - H$, and $H-K_{\rm S}$ of $\sim$6.6\,mag, $\sim$3.4\,mag, and $\sim$1.6\,mag, respectively.
Although further investigation based on infrared photometric and/or spectroscopic studies is necessary for better understanding, if this source is an O star, it would mean that the most massive star in Sh2-48 formed toward the center of the LVC, which corresponds to the center of collision region.
This could be reasonably expected under the CCC scenario, as the effect of gas compression is strongest at the head of the collision.


It has also been suggested that Sh2-48 has evolved only toward the east of the LVC, i.e., filament W. 
If so, this could probably be attributed to the positioning of O stars in the region; formation of O stars in Sh2-48 within the eastern part of the collision region, as shown in Figure\,\ref{fig:ccc4}(b), would excite ionized gas that would be prevented from streaming into the western region by the dense gas in the collision layer, which in turn would have induced the H{\sc ii} region to evolve toward the east.
The asymmetric profiles of the 20\,cm and 24\,$\mu$m emission shown in Figure\,\ref{fig:rotplot} support this assumption. 
To better understand the observed properties of both the neutral and ionized components, it will be necessary to perform numerical simulations to calculates CCC effects including ionization.

\section{Summary}

The conclusions of this study are summarized as follows:
\begin{enumerate}
\item We analyzed the CO $J$=1--0 dataset of the Galactic H{\sc ii} region Sh2-48 obtained as a part of the FUGIN project using the Nobeyama 45-m telescope. Our CO data revealed that the Sh2-48 cloud shows a systematic velocity change over $\sim$5\,km\,s$^{-1}$. In the lower-velocity side ($\sim$42\,km\,s$^{-1}$) the CO emission is spatially coincident with the bright 8\,$\mu$m filament at the western rim of Sh2-48, i.e., filament W, which is elongated nearly perpendicular to the Galactic plane.
As velocity increases to $\sim$47\,km\,s$^{-1}$, the CO emission separates into the eastern and western sides of filament W.  
The position-velocity diagram in the frame perpendicular to the LVC shows that the two components at the higher-velocity range to the east and west of the LVC---the HVC-E and HVC-W, respectively---are connected with the LVC, forming a V-shaped gas distribution in the position-velocity diagram.
\item We quantified the $^{13}$CO opacity of the Sh2-48 cloud by calculating the ratios of $^{13}$CO to C$^{18}$O, and then computed the opacity-corrected $N({\rm H}_2)$ distribution of the cloud. The derived $\tau_{13}$ ranges from 0.5--5.0, with the $N({\rm H}_2)$ estimated as $(4$--$6)\times10^{22}$\,cm$^{-2}$. 
The total molecular mass of the Sh2-48 cloud are $\sim3.8\times10^5$\,$M_\odot$.
\item Comparison of the CO emission with the absorption in H$\alpha$ revealed that the LVC is located in front of Sh2-48, while the HVC-E is located inside or at the rear; nevertheless,  these components are both physically associated with Sh2-48, as indicated in the comparisons with the 8\,$\mu$ and 20\,cm radio continuum maps.
\item 
The location of the right side of the V-shaped velocity distribution, which corresponds to the HVC-W, outside the H{\sc ii} region, along with the lack of an observational signature that the HVC-W interacts with Sh2-48 or other nearby H{\sc ii} regions, make it difficult to interpret the V-shaped gas distribution observed in the position-velocity diagram as expansion of Sh2-48.
We therefore postulated an alternative source of the V-shaped gas distribution in the form of a collision between two clouds. 
We compared the present observational signatures of the Sh2-48 cloud with a numerical model of CCC, and found that, if we assume a collision between a vertically elongated cylindrical cloud and a spherical cloud, the modeled velocity structures observed from an angle nearly parallel to the collision axis  reflect the observations. 
\item We tested an H{\sc ii} region expansion model to estimate the formation timescale of the H{\sc ii} region in Sh2-48 as $\sim$1.3--3\,Myr by measuring the Lyman continuum photons ($N_{\rm Ly}$) of the H{\sc ii} region from the 20\,cm data. 
The derived figure of 1.3\,Myr is consistent with the scenario in which CCC in Sh2-48 triggered O star formation in this region. 
\item The measured $N_{\rm Ly}$ is as large as $\sim10^{48.92}$\,photons\,s$^{-1}$, which corresponds to a single O6.5 star. As only one O9.5 star, BD-14\,5014, has been identified in Sh2-48 to date, we postulated that there are one or more hidden O stars probably located toward filament W, where the peaks of the 8\,$\mu$m, 24\,$\mu$m, and 20\,cm emission are concentrated within a few pc.
This region corresponds to the head of the collision, lending further support to the proposed CCC scenario.
\end{enumerate}

\begin{ack}
The authors thank the anonymous referees for the helpful comments. 
This work was financially supported by Grants-in-Aid for Scientific Research (KAKENHI) of the Japanese society for the Promotion of Science (JSPS; grant numbers 15H05694, 15K17607, 24224005, 26247026, 25287035, and 23540277). 
This work is based in part on observations made with the {\it Spitzer} Space Telescope, which is operated by the Jet Propulsion Laboratory, California Institute of Technology under a contract with NASA. This research has made use of data obtained from the SuperCOSMOS Science Archive, prepared and hosted by the Wide Field Astronomy Unit, Institute for Astronomy, University of Edinburgh, which is funded by the UK Science and Technology Facilities Council. TJH is funded by an Imperial College Junior Research Fellowship.
\end{ack}


\appendix
\section{Velocity channel map of the C$^{18}$O emission}
\begin{figure}
 \begin{center}
  \includegraphics[width=17cm]{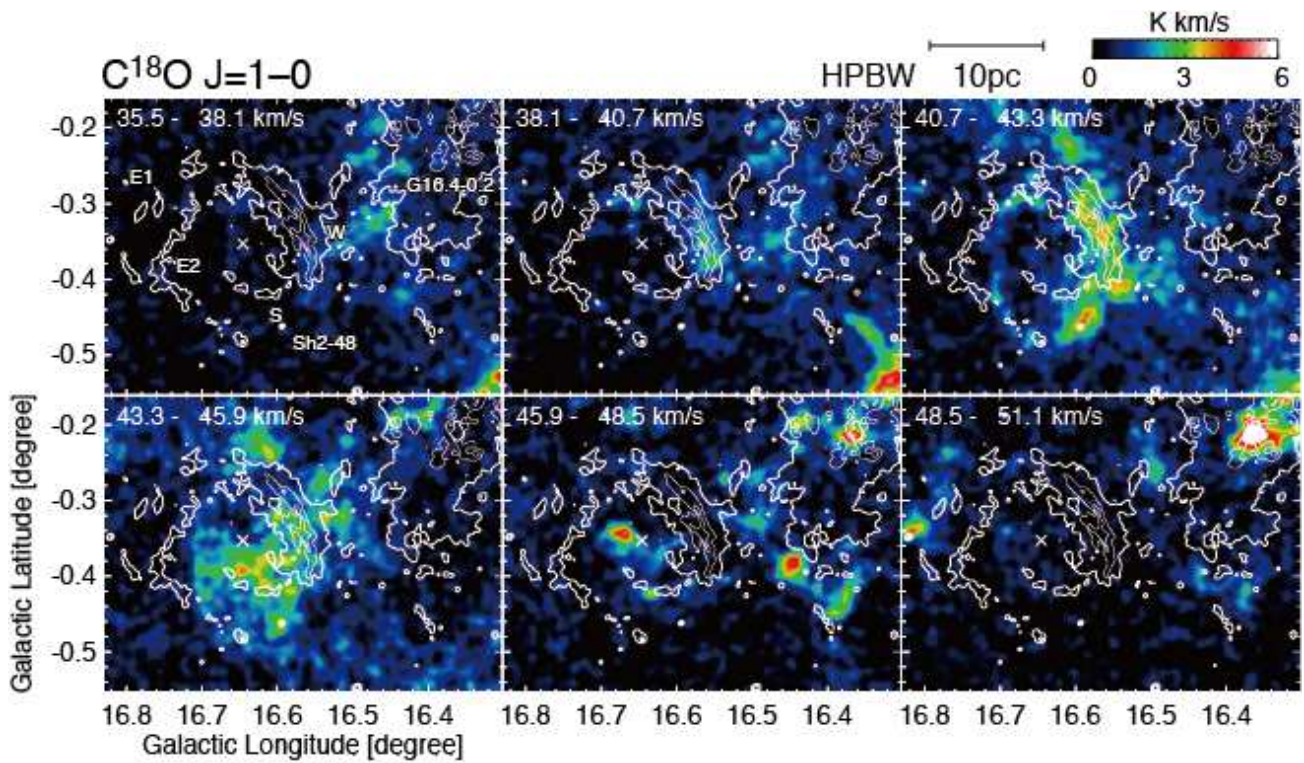}
 \end{center}
 \caption{Velocity channel map of the C$^{18}$O emission in Sh2-48. The cross and the contours are the same as those in Figure\,\ref{fig:ii}. The cross indicates the position of BD-14\,5014, while the contours show the 8\,$\mu$m emission plotted at the same levels as in Figure\,\ref{fig:rgb}(a).}\label{fig:3x2_18}
\end{figure}

Figure\,\ref{fig:3x2_18} shows the velocity channel distributions of the C$^{18}$O $J$=1--0 emission in the same manner as in Figures\,\ref{fig:3x2_12} and \ref{fig:3x2_13}.

\end{document}